\documentclass[11pt,twoside]{article}
\usepackage{asp2014}

\aspSuppressVolSlug
\resetcounters

\begin{document}

\title{RCSEDv2: the largest database of galaxy properties from a homogeneously processed multi-wavelength dataset}

\author{Igor~Chilingarian$^{1,2}$, 
Sviatoslav~Borisov$^{3,2}$,
Vladimir~Goradzhanov$^{2,4}$,
Kirill~Grishin$^{5,2}$,
Anastasia~Kasparova$^{2}$,
Ivan~Katkov$^{6,2}$, 
Vladislav~Klochkov$^{2,4}$,
Evgenii~Rubtsov$^{2,4}$, and
Victoria~Toptun$^{2,4}$}
\affil{$^1$Center for Astrophysics -- Harvard and Smithsonian, Cambridge, MA, USA; \email{igor.chilingarian@cfa.harvard.edu}}
\affil{$^2$Sternberg Astronomical Institute, Moscow State University, Moscow, Russia}
\affil{$^3$Department of Astronomy, University of Geneva, Versoix, Switzerland}
\affil{$^4$Department of Physics, Moscow State University, Moscow, Russia}
\affil{$^5$Universit\'e de Paris, CNRS, Astroparticule et Cosmologie, F-75013 Paris, France}
\affil{$^6$New York University Abu Dhabi, Abu Dhabi, UAE}

\paperauthor{Igor~Chilingarian}{igor.chilingarian@cfa.harvard.edu}{ORCID}{Center for Astrophysics -- Harvard and Smithsonian / Sternberg Astronomical Institute}{}{Cambridge}{MA}{02138}{USA}
\paperauthor{Sviatoslav Borisov}{sb.borisov@voxastro.org}{0000-0002-2516-9000}{University of Geneva}{Department of Astronomy}{Geneva}{}{}{Switzerland}
\paperauthor{Vladimir Goradzhanov}{goradzhanov.vs17@physics.msu.ru}{0000-0002-2550-2520}{Sternberg Astronomical Institute, Lomonosov Moscow State University}{}{Moscow}{}{119234}{Russia}
\paperauthor{Kirill Grishin}{kirillg6@gmail.com}{0000-0003-3255-7340}{Sternberg Astronomical Institute, Lomonosov Moscow State University}{}{Moscow}{}{119234}{Russia}
\paperauthor{Anastasia Kasparova}{anastasya.kasparova@gmail.com}{0000-0002-1091-5146}{Sternberg Astronomical Institute, Lomonosov Moscow State University}{}{Moscow}{}{119234}{Russia}
\paperauthor{Ivan Katkov}{katkov.ivan@gmail.com}{0000-0002-6425-6879}{NYU Abu Dhabi}{Center for Astro, Particle, and Planetary Physics}{Abu Dhabi}{}{129188}{UAE}
\paperauthor{Vladislav Klochkov}{vladislavk4481@gmail.com}{0000-0003-3095-8933}{M.V. Lomonosov Moscow State University}{Department of Physics}{Moscow}{}{119991}{Russia}
\paperauthor{Evgenii Rubtsov}{rubtsov602@gmail.com}{0000-0001-8427-0240}{Sternberg Astronomical Institute, Lomonosov Moscow State University}{}{Moscow}{}{119234}{Russia}
\paperauthor{Victoria Toptun}{victoria.toptun@voxastro.org}{0000-0003-3599-3877}{Sternberg Astronomical Institute, Lomonosov Moscow State University}{}{Moscow}{}{119234}{Russia}



  
\begin{abstract}
The Reference Catalog of Spectral Energy Distributions of 800,000 galaxies (RCSED) includes the results of uniform re-processing of 800,000 SDSS DR7 galaxies at redshifts $0.007<z<0.6$ complemented with ultraviolet-to-infrared photometric data from GALEX, SDSS, and UKIDSS. The key difference between RCSED and existing databases of galaxy properties (NED, HyperLeda, part of SIMBAD) is that rather than providing a compilation of literature data, we perform homogeneous data analysis of spectral and photometric data using our own tools and publish derived physical properties of galaxies along with re-calibrated spectra and photometry and their best-fitting models. Here we present the 2nd release of our catalog, RCSEDv2 where we substantially expanded the spectral dataset to 4 million objects by including spectral data analysis for 10 large spectroscopic surveys (SDSS, SDSS/eBOSS, LAMOST, Hectospec, CfA redshift surveys, 2dFGRS, 6dFGS, DEEP2/3, WiggleZ). The photometric part has also been expanded by including DESI Legacy Survey, DES, UHS, ESO Public Surveys, and WISE in addition to GALEX, SDSS, and UKIDSS used in the original RCSED. This makes RCSEDv2 the largest database of galaxy properties and homogeneously processed spectral and photometric data up-to-date and creates a foundation for the analysis of future large-scale spectral surveys DESI and 4MOST.
\end{abstract}

\section{Motivation and Scope}

Statistical studies of galaxy formation and evolution require knowledge of galaxy properties going beyond simple redshift measurements. Millions of optical spectra became available over the past 2 decades as a result of wide-field spectroscopic surveys carried out at different observatories all over the world. In 2010--2017 we undertook a major effort to re-process 800,000 galaxy spectra from the legacy spectroscopic sample of SDSS. We complemented spectra with aperture and integrated photometry in the ultraviolet wavelengths from Galaxy Evolution eXplorer \citep{2005ApJ...619L...1M}, optical domain from SDSS, and 4 near-infrared bands from UKIDSS \citep{2007MNRAS.379.1599L} converted into rest-frame using $k$-corrections \citep{2010MNRAS.405.1409C} to allow one to compare photometric measurements for galaxies at different redshifts. This Reference Catalog of Spectral Energy Distributions (RCSED) of galaxies \citep{2017ApJS..228...14C} has already been used for a number of successful research projects. We decided to substantially expand the catalog by including virtually all publicly available spectral galaxy surveys and data from new deep wide-field imaging surveys.

\section{RCSEDv2: Data and Analysis}

Spectroscopic datasets included in the RCSEDv2 comprise: SDSS DR16 galaxy and QSO specta \citep{2020ApJS..249....3A}, 2dF Galaxy Redshift Survey \citep{2001MNRAS.328.1039C}, 6dF Galaxy Survey \citep{2004MNRAS.355..747J}, DEEP2/3 surveys \citep{2013ApJS..208....5N}, WiggleZ \citep{2012PhRvD..86j3518P}, LAMOST DR7, public spectra from the Hectospec spectrograph, and several CfA redshift surveys from the 60-inch telescope at Fred Lawrence Whipple Observatory, the most notable being a recently released public archive of FAST \citep{2021AJ....161....3M}. The two latter datasets are collections of uniformly reduced pointed observations from the corresponding data archives, e.g. these are not survey datasets even though some subsets of them are indeed galaxy surveys with specific selection functions. All the remaining datasets are galaxy surveys that is datasets where the target selection functions are well defined. Currently, we have a total of 4.6 million galaxy spectra included in RCSEDv2. For every spectrum, we provide aperture-matched UV-to-IR fluxes from several wide-field imaging surveys (GALEX, SDSS, Legacy Surveys, DES, UKIDSS, UHS, ESO Public Surveys, WISE). We also publish extinction- and $k$-corrected integrated fluxes for most galaxies in the RCSEDv2 sample.

We use the {\sc NBursts} full spectrum fitting technique \citep{2007IAUS..241..175C,2007MNRAS.376.1033C} with several grids of stellar population models and, for a subset of spectra where we had available UV and NIR photometric data, we fitted broadband spectral energy distributions simultaneously with spectra using the {\sc NBursts+phot} code \citep{2012IAUS..284...26C}. For every data source we created a pre-processing routine, which converts the original data into the same format, applies flux calibration and correction for telluric absorption (if needed), prepares a map of wavelength-dependent spectral resolution for each spectrum and finally allows us to run our spectral fitting code. The spectrum fitting  returns stellar (absorption-line) and gas (emission-line) kinematics and a star formation history in a parametric form, typically age, metallicity, and $\alpha$-enhancement of stellar population. We use several grids of stellar population models and include the corresponding parameters of the star formation history in RCSEDv2 for every model grid.

\begin{figure}
    \centering
    \includegraphics[width=0.7\hsize]{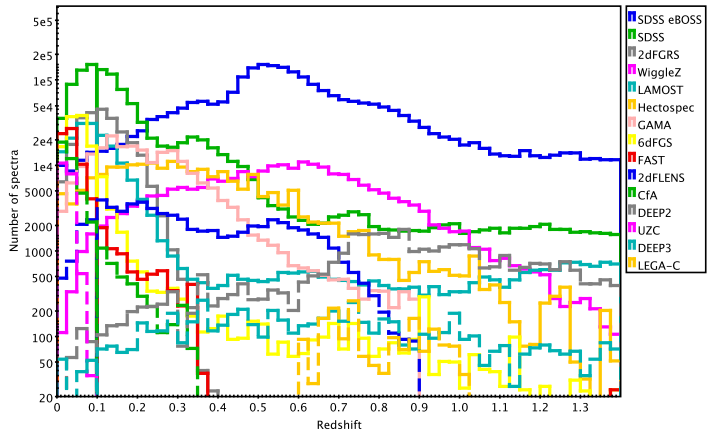}
    \caption{Redshift distributions of RCSEDv2 galaxies originating from different spectroscopic datasets. The $y$ axis scale is logarithmic.}
    \label{fig_redshift_dist}
\end{figure}

\begin{figure}
    \centering
\vskip -5mm
    \includegraphics[width=0.9\hsize]{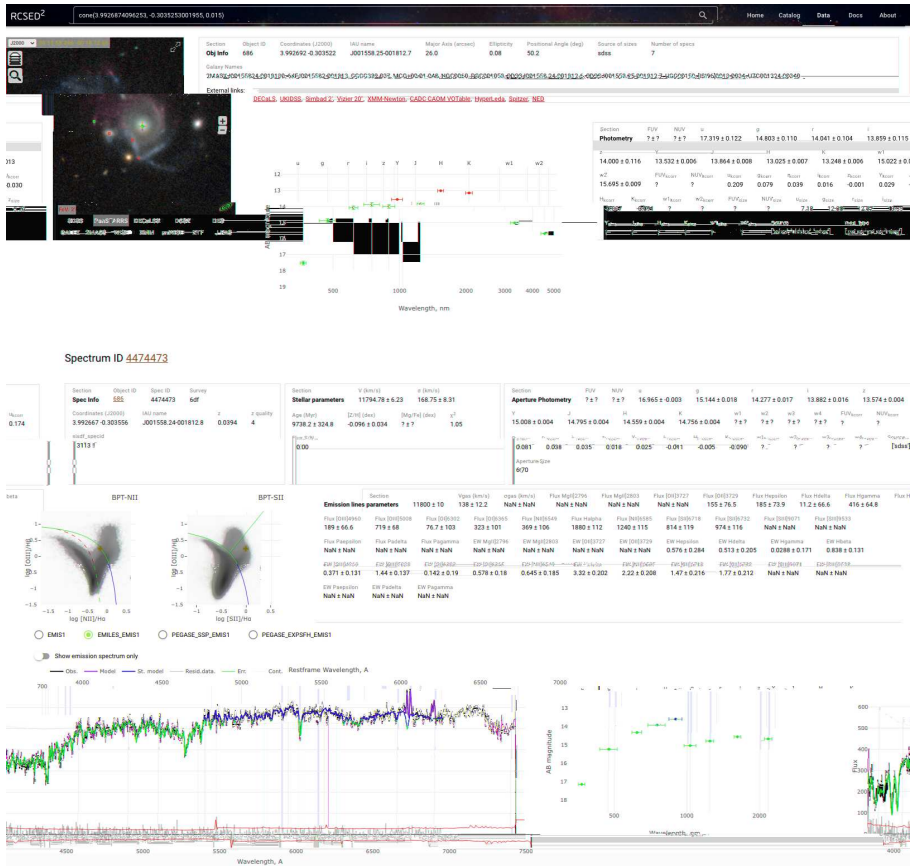}
    \caption{A screenshot of an RCSEDv2 web-page for NGC~60 with a summary of its properties and the first available spectrum with the parameters derived from it. Red markers over a galaxy image display the positions of available spectra.}
    \label{fig_ngc60}
\end{figure}

\section{Technical Solutions and Data Access}

We use {\sc PostgreSQL} database engine with {\sc pgSphere} \citep{2004ASPC..314..225C} as a back-end solution. The intermediate layer for Virtual Observatory access protocols is provided by GaVO DaCHS \citep{2014A&C.....7...27D}. The intermediate-layer uses {\sc Django} for database interaction. The front-end is powered by several modern {\sc javascript} frameworks and also uses {\sc aladin-lite} Sky Atlas \citep{2014ASPC..485..277B}.

\section{Research Highlights and Perspectives}

Using RCSED (i) we discovered the universal optical-UV color-color-magnitude relation of galaxies \citep{2012MNRAS.419.1727C}; (ii) we identified over 70\% of known compact elliptical galaxies \citep{2015Sci...348..418C}; (iii) we found a population of intermediate-mass black holes which power active galactic nuclei \citep{2018ApJ...863....1C}; (iv) we identified progenitors of present-day ultra-diffuse galaxies as low-mass spatially extended post-starburst galaxies in clusters \citep{2021NatAs.tmp..208G}. With RCSEDv2, a new version of our dataset, we have already found young analogs of compact elliptical galaxies, substantially expanded a sample of intermediate-mass black holes, and found many active galactic nuclei in rare giant low-surface brightness galaxies \citep{2021MNRAS.503..830S}. 

Numerous interesting research results in the field of galaxy formation and evolution and in the studies of the large-scale structure of the Universe are expected to be spawned by the analysis of the largest homogeneously processed extragalactic dataset provided by RCSEDv2. We expect to use the infrastructure and data processing and analysis approaches to the data which will be delivered by the next-generation large spectroscopic surveys DESI and 4MOST.

\acknowledgements This project is supported by the Russian Science Foundation Grant 19-12-00281 the Interdisciplinary Scientific and Educational School of Moscow University ``Fundamental and Applied Space Research''.



\bibliography{X3-008}


\end{document}